\documentclass[12pt]{article}
\usepackage[cp1251]{inputenc}
\usepackage[T1]{fontenc}
\usepackage{indentfirst}
\usepackage[dvips]{graphicx}
\begin{document}

\title{Single- and Multichannel Wave Bending}

\author{B.~N. Zakhariev and  V.~M. Chabanov \\
Joint Institute for Nuclear Research,       \\
Laboratory of theoretical physics, \\
141980 Dubna, Russia,  \\
e-mail :zakharev@theor.jinr.ru; URL : http:// theor.jinr.ru/~zakharev/}
\date{} \maketitle

\begin{abstract}
It was an important examination to  give a review
talk  at  the previous  Conference on  Inverse Quantum Scattering
(1996, Lake Balaton) about computer  visualization of this science
in front of  its fathers- creators ,  B.M. Levitan and V.A.
Marchenko. We have achieved  a new understanding that the
discovered main rules of transformations of  a single wave
function bump, e.g., for  the ground bound states of one
dimensional
 quantum systems  are applicable to any state of any potential with arbitrary number
 of bumps from finite  to unlimited ones as scattering states and bound states embedded
into continuum. It appeared that we need only to repeat the rule mentally the necessary
 number of times.  That uttermost simplification and unification of physical  notion
of spectral, scattering and decay control for  any potential have got an obligatory
praise from B.M. Levitan at the conference and was a mighty stimulus for our further
research After that we have written  both Russian (2002) and improved English editions of
 “Submissive Quantum Mechanics. New Status  of  the Theory in Inverse Problem Approach”
\cite{1} (appeared at the very end of 2007).  This book was written
 for   correction of the present defect in quantum education throughout the world.
 Recently  the quantum IP intuition helped us to discover  a new concept of  permanent
 wave resonance with potential spatial oscillations \cite{2}.  This means the {\bf constant
  wave swinging frequency on the whole energy  intervals} of spectral forbidden zones
 destroying physical  solutions   and deepening  the
theory of waves in periodic potentials. It  also shows the other
side of strengthening the fundamentally important magic
structures.  A ’new language’ of {\bf wave bending}
 will be presented to enrich our quantum intuition, e.g., the paradoxical effective
  attraction of barriers and repulsion of   wells in multichannel systems,
  etc.
\end{abstract}


\section{ Introduction}
The inverse problem (IP)  infinite number of exact models instead of only about
ten previously known were  a great present of our mathematicians (Gelfand-Levitan-Marchenko)
to quantum physics. But they were for a long time not enough understood and used.
Here will be helpful the discovery of qualitative golden  rules of transformations
of the most elementary wave "bricks", their separate bumps and the notion about the
corresponding simplest and fundamental building blocks of potentials . This opened the
'black box' of computer visualization of these models and intensifies many times their
evidence and usefulness. It means a real quantum ABC to acquire the quantum literacy and
facilitate the future discoveries. So, one gains the absolutely unexpected ability of
immediate prediction how, in principle, to achieve the precise and evident the spectral,
scattering and decay control, particularly for multi-channel systems to obtain the
given properties of the constructed objects. An unprecedented combination of qualitative
simplicity and clarity with absolute exactness was achieved. That was impossible
to imagine in the previous quantum theory. Unlike the numerous books on quantum
mechanics, mainly compilations, the one  with approach from IP \cite{1} has no analogs in
contemporary  literature. It is utmost intelligible due to computer visualizations. It
is not a substitution of the traditional books, but a
fundamental, strengthening and simplifying addition to them enlarging and deepening the
understanding of the subject instead of previously unintuitive and approximate approach.
Investigation of exact solvable models allowed us to reveal the possibility of  practically
infinite simplification of  qualitative  prediction “in mind” of continuous manifolds
of  elementary transformations of  quantum systems by variation of separate spectral
parameters (energy levels, spectral weights, etc.).   See references to  independent
opinions on this theory in Internet on the site http://theor.jinr.ru/$\sim$zakharev/.

Among some wonderful, instructive examples of  wave behavior found
by us  in IP approach
and explained can be mentioned  the following \cite{3} ones.  Seemingly improbable
{\bf coexistence of confinement and unlimited} propagating waves
as possible
solution of the same quantum system at the same energy \cite{3}, 1999. Unusual (non-Gamow)
decay states can be constructed by shifting energy levels into the complex  $E$
plane \cite{3}, 2001. Resonance absolute quantum reflection at selected energies was shown
the first time  and on exactly solvable IP  model \cite{3} , 2001. The surprising equality
of  transparencies of separate potential parts being cut in arbitrary point (one- and
multichannel cases \cite{3}, 2006.  Unusual  discretization  of scattering states \cite{3}, 2006.

The experience of consideration of computer visualization of continuum transformation
exactly solvable inverse problem models of quantum mechanics   helps us to imagine
{\bf the rules} of ‘motions’ in corresponding infinite dimensional exact model spaces.
This possibility we have got instead of consideration ‘separate points’ of rectangular,
oscillator and other previously known rare analytically solvable  examples. This will
be a step to more perfect feeling the essence of micro-world's  wave mechanics. Here
for human brain appears to be accessible what is beyond the possibilities of numerical
calculations even for most perfect modern and may be also for future computers.

We emphasize the notion of the strongest sensitivity of the waves $V(x)$   in
neighborhood of bumps extreme points (of maximum spatial density).  So IP by
revealing the essence of connections: spectral weight with potential and quantum wave
shape gave
us the main hint to one-channel and later (recently) the multi-channel wave bending.
It was crucial for formulation  rules of spectral transformations:  control of wave
localization and energy level shifts. And recently as a remarkable illustration
it served the deepening of understanding the mechanism of forbidden zone formation \cite{2}.

It would be useful to mention that our experience of quantum intuition,
which we got considering thousands of solutions exactly solvable models (ESM)
we have spread to simple predictions, which became evident, for much more broad
set (manifold) of problems, not ESM, without requirement to keep unchanged
all spectral parameters except chosen ones what strongly enriched the possibilities of
{\bf direct} problem “solutions in mind”. Really, the new language of wave bending based
originally on the inverse problem approach, although after mastering these notions
it became not absolutely necessary to connect them with the inverse problem. There
already works successfully the simple quantum logic.  It is especially so for extension
of the qualitative  theory to coupled equations , e. g.
multi-channel systems [1,2]. It reveals a lot of physical effects,
which can be expected due to the more powerful formalism possible
to take into account inner degrees of freedom of investigated
objects, e.g., the unexpected inversion of usual asymptotic
behavior  of partial channel waves with different thresholds of
continuum spectral branches (in less closed channel functions can
decrease more rapidly).
Unfortunately the main part of the physical community is still unfamiliar with the
new aspects of quantum  literacy and the “intuitive spectral theory”
which successfully continues  its development.

\section{Rules of Bending Solutions of 1D Stationary Schr\"o\-din\-ger Equation and
Exact Inverse Problem Models}

To solve Schr\"odinger equation
\begin{eqnarray}
-\Psi''(x) = (E-V(x))\Psi  (x)
\label{Schreq}
\end{eqnarray}

the suitable program is inserted into computer as into
the “black box”, which gives at the exit some numerical results.
This is not enough to create the clear notion of how the shape of $V(x)$ determines the
behavior of $\Psi  (x)$ even if we get also the graphical illustrations. But generally
the science needed so much to rise the prestige of clarity for accelerating its
progress. Now it is rather widely spread among some authors the practice to hide behind
the scientific-like fog the lack of understanding the subject of their own research.

Let us mention the simple, but very important fact which was left till now without
necessary attention. Equation (\ref{Schreq})
shows how the intensity of wave function $\Psi  (x)$ bending, its second derivative
$ d^{2}\Psi (x)/dx^{2}$, is determined by the potential $V(x)$. The degree of $V(x)$
influence in different space points $x$ is proportional to the factor $\Psi  (x)$,
with which $V(x)$ enters into the(\ref{Schreq}). And this factor $\Psi (x)$
when oscillating has its absolute value between zero at wave knots and maximums
at its extreme values. This changes the sensitivity of $\Psi (x)$  to  $V(x)$, so that
$\Psi (x)$ takes part directly in control of its own bendings $ d^{2}\Psi (x)/dx^{2}$.
We ourselves for a long time have missed the simple rules of  bending $\Psi (x)$
before revealing that: {\bf for positive local kinetic energy $E-V(x)$
the real wave is bending TO $x$ axis independent to the sign of wave}. For negative
$E-V(x)$ the bending  take place in opposite direction: {\bf  from the $x$-axis}.
In the case of {\bf free } motion  $V(x)=0$ this rule gives the obvious
notion of the solution, which in
this simplest case the well known  from numerical calculations are illustrated  by Fig.1

The qualitative pictures, see Fig.1, convincingly illustrate the given bending elementary
rules to facilitate “solutions in mind”.

\begin{figure}
\begin{center}
\includegraphics[height=6cm,keepaspectratio]{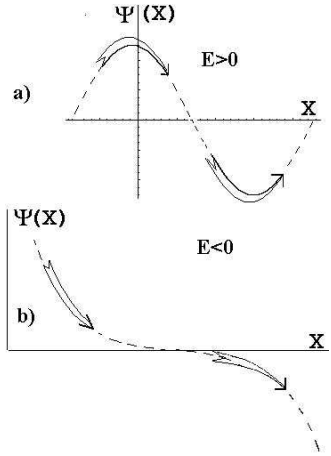}
\end{center}
\caption{{\small Bending of wave function when a) kinetic energy is positive b) kinetic
energy is negative}}
\end{figure}
This determines the local change of solution and its derivative. When the
function approaches to $x$ axis its bending decreases and at the
knot point disappears at all (converges to linear line). After the knot the sign of the
wave function changes and it gets an opposite bending. But as {\bf before the knot}, the
bend continues to be {\bf TO} the $x$ axis. Further this approach will facilitate such
considerations in more complicated cases {\bf of nonzero and even matrix potentials}.

An interesting example of wave bending is the case of perturbation of initial
free motion with real sine-like waves by periodic potential $V(x)$. Here we
revealed seemingly paradoxical conception of the {\bf permanent resonance}.
In each point of the separate continuous energy intervals $[E_{<},E_{>}]$
there is exact coincidence of commensurability of spatial oscillations of
{\bf fixed (!)} potential and the strongly {\bf energy dependent (!)}
solution $\Psi (x)$ \cite{3}. This was understood due to inverse problem approach
\cite{3}. A remarkable lesson one gets here consists in the following. The
solutions of Schr\"odinger equation can “feel” {\bf the same fixed}
periodical potential sometimes as attractive, but sometimes as repulsive.
It depends on {\bf the shift of wave chain relative the potential lattice}.
These shifts change the effective strength of potential influence on wave
motion. The most repulsive effective  potential becomes when wave bumps
embrace by their main sensitive regions the repulsive zones of potentials.
Analogous situation is for the most attractive effective potential when
wave bumps embrace by their main sensitive regions the attractive
potential zones. Both these cases are achieved simultaneously with the
symmetry of wave bump derivatives at the knots at different energy values
$E_{<}\, < \, E_{>}]$, see Fig.2a,b. This depends on shifts of solution bumps (by special
choice of boundary condition, e.g., fixing the position of some wave knot)
relative to barriers and wells of periodical potential. The potential
has in general {\bf increased repulsion} when barriers are posed in
vicinity of wave density (bumps absolute value) maximums,  susceptible
“wave tentacles”, and wells in vicinity of knots, “less sensitive wave zones”.
The “attractiveness” of $V(x)$ is increased  when potential wells are near
the extrema of wave bumps and barriers are in week sensitive neighborhood
of knots.  The changes of $V(x)$ averaged over the shifted period allow to
compensate the total energy $E$ changes. So that with increase the energy
the averaged potential increased by the same value (with suitable space
shift of solution) keeping {\bf constant the averaged kinetic energy $(E-V)$,
and the frequency of wave oscillations}.  In other words, the mutually
coordinated x- and
E-shifts of solutions allow to compensate the changes
of the full energy $E$ by corresponding potential variations to “freeze”
the effective kinetic energy $E_{kin}$. This allows
{\bf “stretching”  exact resonance at one point into permanent one on the whole
forbidden zone}. This explains the wonderful
exact coherence,  of waves and periodical potentials destroying the
physical solutions over the whole energy intervals of forbidden zones
(spectral lacunas) \cite{3} 2004. Another behavior has the usual  free real wave. Its
{\bf space oscillation frequency is increased with  any small rise of energy}.
Here the shift of one knot moves the whole solution over the $x$ axis with exact
conservation of its form.

\begin{figure}
\begin{center}
\includegraphics[height=6cm,keepaspectratio]{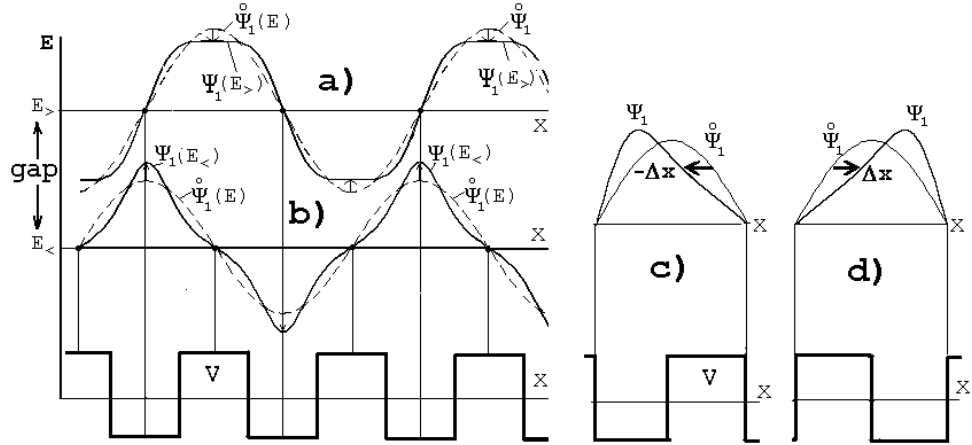}
\end{center}
\caption{{\small a)Bending of wave function $\Psi_{1}(E_{>})$
which is  shifted so that its bumps embrace symmetrically
potential barriers and feel periodical potential as effectively
strong repulsive. b) The bumps of $\Psi_{1}(E_{<})$  embrace
symmetrically potential wells what is acting as average effective
strong attraction. c,d) Bumps of $\Psi_{1}(E)$ under which  are
disposed well-barrier or barrier-well potential blocks distort the
symmetry of wave bumps, see figures of x-shifts in [1].. The
crucial fact is here the equality of bump lengths at different
energies due to equality of averaged over the bumps kinetic
energies $(E-V(x)\Psi(x)$. At all values of the full energy $E$ on
the whole energy interval $[E_{<},E_{>}]$ the average kinetic
energy is constant. It is due to compensation of $E$ variations by
effective average potential for different space shifts of wave
bumps (knots)relative to the potential lattice. The asymmetric
bumps cause the exponential growth of amplitude of wave swinging
without changing knot's positions (smooth continuation of bumps
requires increase/decrease the neighbor bump by the same factor on
each period resulting in exponential growth of amplitude of wave
swinging. So the conservation of bump lengths provides the exact
permanent wave-potential resonance increasing waves and destroying
the physical solutions on the whole forbidden zone
$[E_{<},E_{>}]$}}
\end{figure}

The periodic solutions as in the cases shown in Fig2a,b are
possible  only  at the {\bf spectral zone boundaries}. There are
no other periodic wave functions (‘paradox’ of periodic
structures)!
Toward the both sides of these boundaries there is violation of the above
mentioned symmetry of derivatives. Only in the region of exact resonance  between
$E_{>}$ and  $ E_{<}$ it occurs with increasing the amplitude
of swinging by the same factor
on the period: the destruction of physical solution (forbidden zone). This leads
to the exponential increasing (or decreasing) of amplitude of wave oscillations
by distorting  bumps, see Fig.2c,d, Fig.3.
Remarkable is the fact that the degree of this distortion and of
forbiddenness at the chosen energy point of the spectral gap are
under control in inverse problem  approach \cite{1}. In addition
infinite number IP of  exactly solvable periodic models can be now
used instead of only two (Kronig-Penny and Dirac combs of delta
potential peaks or wells).

And {\bf from the other side of the spectral gap boundary the
resonance is approximate} and with accumulation of phase shifts of
potential lattice and wave space oscillations there happens the
alteration of regimes of solution growth from one direction to the
opposite direction. This results in “healing the tumor like
expanding”: the  swinging  decreases compensating the growth and
then again begins the growth and so on providing formation {\bf
beating } of solutions in allowed zone. The beating remind
something  the periodical modulations and locally the  processes
increase/decrease  in forbidden
 zone, but with complete mutual compensation of dangerous growth in opposite directions.

In forbidden zone we get two linearly independent exponentially
swinging with increasing amplitude in one or other side, see Fig.3. They are especially
convenient due to their simplicity  with evident properties for using as fundamental
solutions: for construction arbitrary other solutions as their linear combinations.
If we shift the origin into one knot of this solutions and cut there the periodic
structure by infinite potential barrier, then it can be considered as example of the
surface (Tamm’s) bound states \cite{4}  on the half
axis $x$ in the direction of decreasing, see Fig.3.

\begin{figure}        \begin{center}
\includegraphics[height=7cm,keepaspectratio]{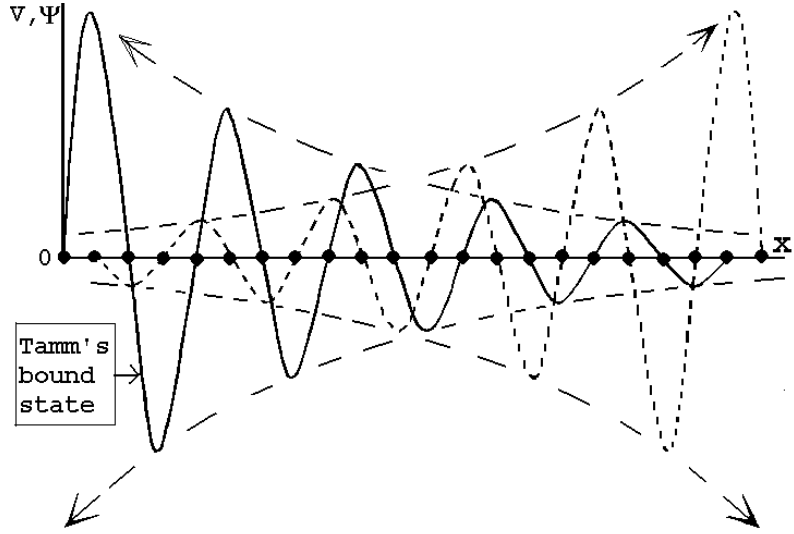}\end{center}
\caption{{\small Two linearly independent fundamental solutions
swinging with exponentially increasing/decreasing amplitudes
(corresponding to Fig.2c/d). They have {\bf equidistant
alternating} knots. The period of knots coincides with the period
of the potential perturbation.  At the energies of zone boundaries
these solutions oscillate with the constant amplitudes and belong
to both the forbidden and allowed zones simultaneously. Tamm's
surface state is obtained by the cut of $x$-axis at the origin at
the energy value $E$ when one of the knots of solution decreasing
to the right hand side is at $x=0$ and there is not penetrable
potential wall }}\label{smart}
\end{figure}

Such {\bf physical state on the half axis
constructed of not physical solution on the whole axis} can be only at one energy
point  for the whole forbidden zone, because all other not normalizable solutions
have no knots at the origin \cite{2}.

It is natural also to expect that the exact rules of spectral
control, found in single-dimensional case, must have some common
features in more general situation. Of course, the most sensitive
parts of wave functions in  multidimensional space will also
concentrate in vicinity of wave density maximums. And shifting of
energy levels must require corresponding potential barriers
(wells) disposed in these regions and
compensating wells (barriers) along the knot-lines. But it is still not clear,
whether it could be  achieved as in more than one dimension.

\section{New Conception of Magic Solidifying of Quantum Systems and  the Principle of Minimal
 Action Promoting the Corresponding Transformations }

Periodic "hammering" by alternating potential barriers and wells distort the
symmetry of wave bumps. They are crumpled during gathering in
one or opposite direction. This results  in exponential increase (decrease) of swinging
amplitude of space wave oscillations at smooth continuation of one bump over the whole
axis. This destroys the normalized physical solutions. It happens inside forbidden spectral
zones in accordance with new conception of {\bf permanent}
resonance mechanism. This special conserving the necessary precise
space oscillation frequency gives resonance having maximum {\bf
not at one energy point}, but on the {\bf whole continuous energy
intervals}, spectral gaps. From one side the   nature of band
spectrum of periodic structures  has some common features with
destroying of bridge by periodic perturbation (wind, soldier's
lockstep) with frequency coincident with eigen-frequencies
 of the bridge oscillations). From the other side it may seem
"paradoxical" that despite this  strong perturbing shaking blows
the  periodic structures  (crystals, chain molecules and polymers)
become strengthened(!). This is a  simply resolved 'paradox' : the
created spectral lacunae prevent some
internal excitations of crystal structure restricting the electron transitions between
allowed zones. It is like the ferro-concrete is reinforced by metallic carcass which hinders
deformations.  It appears that analogous principle operates in cyclic wave
motions forcing apart bound state energy levels in {\it magic} atomic and nuclear
constructions, organic ring molecules, nanotubes, fullerenes. This
new language  allows to unify main aspects of seemingly quite
different bound states: in forbidden zones and ones embedded into
continuum as also creation and removal of bound states, of
resonance and quasi-bound states control and so on \cite{1}.

\subsection{Hypothesis of self regularization}
The fact of  strengthening the periodic structures by means of  spectral gaps was a hint to
the new idea about possibility of reconstruction of  forbidden zones under addition of
impurities into crystals.  It is right that the chaotic additions  (admixtures)  destroy
periodicity and the resonance mechanism of creation of forbidden
zones.  But the energy gain which gives the periodicity
(especially with filled zones),  promotes the  rebuilding the
system. According to the principle of the  minimum action  it is
advantageous  to the system to aspire to increase of  the binding
energy. The particles of admixture endeavor to  distribute
themselves so that there appears new periodicity and together
with it the new band spectrum. So, in seemingly unpredictable conditions of accidental
impurities  suddenly was clarified their tendency to their possible self regularization.

\subsection{New glance at the property to create magic structures in atomic, nuclear systems,
molecular rings, chains, polimer, nanotubes, fullerenes}

In cyclic wave motions the analogous destruction of physical solutions
as in periodic systems occurs due to their not smooth sewing  at the points of beginning
and end of the cycle (instead of swinging with increasing/decreasing  amplitudes).
The corresponding spectral gap between the energy levels of bound
cyclic states as in crystals, strengthens the cyclic systems. It
allows a new look at the effect of magic solidity. It is true,
that in one-dimensional case the allowed zone consist of a single
energy level. Here the boundaries of allowed zones are as if
sticked together into one energy level. In multi-dimensional,
multi-channel systems the number of allowed states increses
without the great changing the essence of the mechanism of the gap
creation.  Here as in periodic case the moving apart of levels,
the widening of spectral gap happens due to increase of potential
variations  on the cycle. So, e.g. deformation of nuclei, if it
makes deeper the effective potential
wells and higher the barriers on the cycle motion of clusters or separate particles,
can assist the appearance of  magic structure. This may be the cause of the stability
of recently discovered  uneven isotope of Al. It may happen if such mutual potential  blows
of the system  constituents prevail (overcome) here the usual
mechanism of strengthening nuclei.

\section{Systems of  Schr\"odinger Equations. Unexpected Role of Nondiagonal
Elements of Interaction Matrix in Bending of Partial Channel Wave
Functions }

Generalized and instructive rules of bending multi-channel
partial waves $\Psi _{\alpha }$ were recently found. They
open the way to better understanding  multi-dimensional and many-body systems. The
formalism of vector functions with partial components $\Psi_{\alpha}(x) $ and matrix
interaction $||V_{\alpha \beta }(x)||$ instead of
scalar potential  $V(x)$ was considered by us in \cite{1} 2000, see also references therein.
 For simplicity we shall consider here the two-channel case
\begin{eqnarray}
-\Psi''_{1}(x) =(E_{1}- V_{11}(x))\Psi_{1}(x) -V_{12}(x)\Psi_{2}(x) \nonumber \\
-\Psi''_{2}(x) =(E_{2}- V_{22}(x))\Psi_{2}(x)-V_{21}(x)\Psi_{1}(x),
 \label{system2}
\end{eqnarray}
where $E_{\alpha }=E-\epsilon_{\alpha }$ and  $\epsilon_{\alpha}$
are threshold values of continuous spectra  of different channels
spectral branches.
The qualitative results here will be valid in principle for general
system with  $ V_{\alpha \beta }(x)$  coupling of arbitrary $\alpha $
and $_\beta  $ channels.
That is true, the diagonal elements   $V_{\alpha \alpha }$  of
{\it interaction matrix} influence on   $\Psi _{\alpha }$ like the
scalar one-channel potentials. Only $V_{\alpha \alpha }$
constitute  the minority among all other matrix elements.  But
nondiagonal ones seemed as some chaos of interactions. And
suddenly it appears, that situation is rather simple. Almost as
for scalar case if $\Psi_{\alpha}(x)$ has the same sign as
$\Psi_{\beta}(x)$. The opposite situation is if there are
different signs  of waves in corresponding channels
at the given point $x$. Here is possible the unexpected {\bf inversion} of
influence their barriers and wells. Namely, for channel wave functions
$\Psi_{1}(x)$  and $\Psi_{2}(x)$  with opposite  signs
{\bf "barriers can be attractive(!)"} increasing the local kinetic energy and {\bf "wells
repulsive (!)"} reducing it. Really $V_{12}(x)$  in the term
$V_{12}(x) \Psi_{2}(x) $  in equation of the first channel in the
system  \ref{system2}  acts as potential with opposite sign when
$\Psi_{1}(x) $ and $\Psi_{2}(x) $  have different signs. To
explain  better let us write this term in the form
$V_{12}(x) \frac{\Psi_{2}(x)}{\Psi_{1}(x)}  \Psi_{1}(x)  $
(multiplying and  dividing by the same function $\Psi_{1}(x)$)  as if with
{\bf  effective
single-channel (scalar) potential}  $V_{12}(x) [\Psi_{2}(x)/ \Psi_{1}(x)] $,
which has opposite sign relative to   $V_{12}(x)$  because of
negative sign of  $[\Psi_{2}(x)/ \Psi_{1}(x)]$.  It means that the
barriers in   $V_{12}(x)$ contribute to bending  $\Psi_{1}(x) $ TO
THE AXIS $x$, and  wells act conversely to turn $\Psi_{1}(x) $
FROM AXIS $x$. So they influence  in opposite manner in comparison
with usual potentials  on the intensity ($-\Psi”_{1}(x) $  of
bending $\Psi_{1}(x) $. Meanwhile in the second channel happens
the analogous inversion.
The corresponding unusual, but nevertheless, simple rules are useful for understanding
some previously mysterious quantum peculiarities. It is illustrated by different
examples, e.g. complex  potentials \cite{4}, periodic structures, transparent interaction
matrices etc.
For similar signs of channel functions  $\Psi _{\alpha }, \, \Psi_{\beta }$
the situation is standard
one: the elements of interaction matrices $V_{\alpha \beta }$ are like the usual
scalar potentials and diagonal matrix elements $V_{\alpha \alpha }$.  Their wells (and
barriers) increase (and decrease)the bending intensity of partial waves to the $x$ axis.
Barriers in $V_{11}(x)$ and $V_{12}(x)$ in both terms $-V_{11}(x))\Psi_{1}(x)$
and $-V_{12}(x)\Psi_{2}(x)$ in the first equation are subtracted
from the energy value $E_{1}$ and make smaller the bending intensity |$\Psi"_{1}(x)$|.
Analogously act the potential matrix elements in the second
equation in \ref{system2}. Wells in $V_{12}(x)$ change
|$\Psi"_{1}(x)$| and |$\Psi"_{2}(x)$| also as in diagonal matrix
elements $V_{11}(x)$ and $V_{22}(x)$.

The consideration of wave bending helps the deeper understanding
of periodicity.  This is an  significant step to solve
qualitatively systems of coupled
equations  'in mind'. So could be explained the mechanism of simultaneous permanent
resonance in channels with different thresholds which  seemingly
violate the necessary equality of "average channel kinetic
energies" on the periods.  But it can be compensated by terms
with  effective  potentials representing the influence of $V_{12}$
(details will be presented elsewhere.),
see also \cite{1} 2000.  We must only mention that the system of two ordinary differential
equations can have two branches of band spectra with partial overlap of forbidden zones.

For understanding the resonance mechanism of spectral gap creation
for waves motion in periodical potentials we used the notion of energy level shifts in
infinite rectangular well when shifted states at different
energy values have the same wave length. Now we
get explanation how the length of different channel waves corresponding to different
thresholds are commensurable due to the action of  of interaction matrix. Although
the zero boundary conditions at the common vertical potential walls of different
channels will automatically provide commensurable wave length of space oscillations,
but it remained unclear, what special role play here separate elements of interaction
matrix. The significance of the notion about the ‘normal’ and ‘inverse’ influence
of $V_{12}(x)$ becomes more apparent if we imagine improbable situation that nobody
knows now trivial fact that the potential barrier is repulsive and well attractive.

\section{Complex Potentials}

Let us now consider the case, see \cite{4}, of complex potential
$V(x)=V_{R}(x)+ i V_{I}(x)$ as an intermediate step to
multi-channel formalism. Here the role of partial channel wave
functions play the real and imaginary wave components $R(x)$ and
$I(x)$:
\begin{eqnarray}
-R''(x) = (E-V_{R}(x))R(x) + V_{I}I(x) \nonumber \\
-I''(x) = (E-V_{R}(x))I(x) - V_{I}R(x). \label{systIm}
\end{eqnarray}

The coupling between equations for real and imaginary components
$R(x),\,I(x)$ of wave function is realized by the imaginary
component $V_{I}(x)$ of the potential $V(x)$. Pay attention to
opposite signs of coupling terms in different equations. So if
$R(x)$ and $I(x)$ have the same signs, {\it the barrier} in
$V_{I}(x)$ in the first partial equation is {\it "paradoxically
attractive"} increasing the bending intensity $-R''(x)$ and
"repulsive" well decreases it. In the second equation $V_{I}(x)$
acts as usual potential. For different signs of $R(x)$ and $I(x)$
the situation is inverted, but again $V_{I}(x)$ acts differently
in both equations. In the case of periodic complex potential the
following  remarkable effect can be clearly understood in terms of
wave bending. The {\it difference in  coupling} of equations
\ref{systIm} hinders the simultaneous resonance in both equations
\ref{systIm} necessary for creating forbidden zones \cite{2} 1997,
2004. So it became quite evident the effect of vanishing gaps and
merging allowed zones when switching on imaginary components in
real periodical potentials perturbation \ref{systIm}.

There are also other
interesting manifestations of bringing to light the essence of
quantum mechanisms. For example unlike the just considered case
how to help permanent simultaneous resonance in coupled partial
channel equations with different thresholds of continuum spectra
which seemingly hinder realization of resonances.

New results are often not understandable for those who needed them. But gradually
we succeed nevertheless in simplification and unifications of our separate notions
and conceptions. Even at the late stages of theory development there suddenly
appeared the laws which seemingly should be revealed long ago.

There was found \cite{1} an interesting example of inversion of channel forbiddingness.
It happens due to the special type of  coupling $V_{12}(x)$. It can be demonstrated
by exactly solvable multichannel model of reflectionless interaction matrix.
Paradoxical is the fact that potential coupling $V_{12}(x \rightarrow -\infty )
\rightarrow 0$ converging asymptotically to zero can pump out the wave from the
first channel with smaller threshold $\varepsilon _{1} <\varepsilon _{2} $. So
the function  $\Psi _{1}(x)$ becomes more rapidly decreasing than $\Psi _{2}(x)$
corresponding to a bigger threshold in contrast to
what was expected. Mention some analogy with the case of BSEC.

A special interesting case is the interaction matrices transparent
at any energy.   Depending on relative signs of $\Psi_{1}(x)$ and
$\Psi_{2}(x)$ in two-channel system with equal thresholds there
will be  free motion or strong perturbation (see Fig.4). In both
cases this perturbation is totally transparent. But in one case it
realises because of cancellation of terms with $V_{11}(x)$ and
$V_{12}(x)$; $V_{22}(x)$ and $V_{12}(x)$. And in other case
because the {\bf halves} of transparent   soliton-like (a)
$V_{11}(x)= V_{22}(x)$ potentials are added so that combine the
whole reflectionless soliton-like transparent effective potential
with well $V_{12}(x)$ (a) or (b) barrier $V_{12}(x)$
potential (with bound states having equal(a) and opposite (b) signs).
The signs of norming factors in $\Psi_{1}(x)$ and $\Psi_{2}(x)$  provide the necessary
results.

\begin{figure}[h]
\begin{center}
\includegraphics[height=8cm,keepaspectratio]{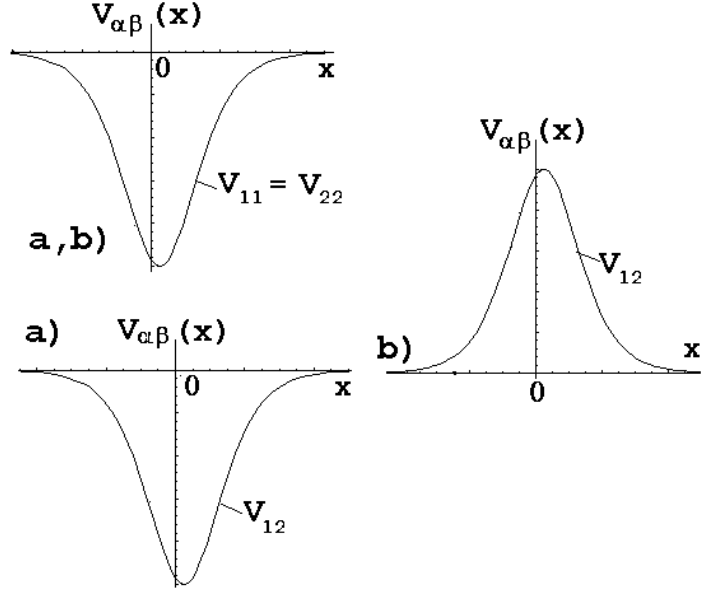}
\end{center} \caption{ {\small Two-channel interaction matrix of transparent
system with equal thresholds and bound state having equal spectral
weights (a) or with opposite signs (b). The shapes of all matrix
elements are equal to the {\it half} of single channel
reflectionless soliton like potential up to sign. The proper
summation of such interactions gives the total transparency of the
system. }} \label{2-ch-solit}
\end{figure}

Let us consider the last instructive example of special role of
nondiagonal singular  interaction matrix elements $V_{12}(x)$.
It is the  construction of two-channel bound state
embedded into continuum (BSEC).

If the function of the single open channel $\Psi_{1}(x)$ nonzero
on the finite interval $[0,a]$ is made to be zero together with
its derivative at the points $0, \pi $ due to the coupling with
other channel  we shall get BSEC because $\Psi_{1}(x<0, x\ge \pi)
\equiv 0$ what is required by {\bf both} zero boundary conditions
at $x=0,\pi $.  The necessary breaks of BSEC-solutions at $x=0,\pi
$ are performed by singular terms with $V_{12}(x=0,\pi)$ and
nonzero $\Psi_{2}(x)$. This is impossible for scalar potentials
and their diagonal analogs at knot points of $\Psi_{1}(x)$.
Fig.5 shows  the BSEC in two channel case with two point delta
interaction matrix totally reflecting waves due to possibility to
break channel solution at the knot so that further it is continued
with zero derivative what cannot be achieved in one channel case.

\begin{figure}[h]
\begin{center}
\includegraphics[height=8cm,keepaspectratio]{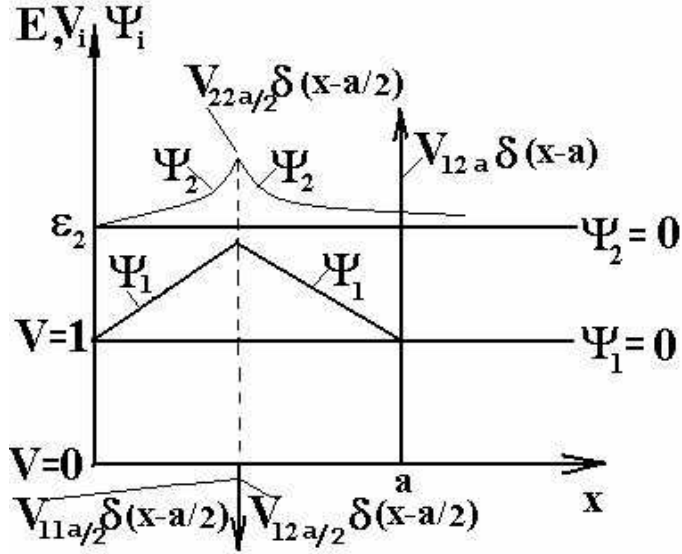}
\end{center} \caption{ {\small Two channel interaction matrix with BSEC
$\Psi _{1}(x), \Psi _{2}(x)$. Although the first channel is open
at BSEC energy in continuous spectrum above $E=\varepsilon _{1}=0$
the partial channel wave $\Psi_{1}(x)$ is zero outside the
interval $[0,\pi ]$  due to its breacking by nondiagonal coupling
of channels. All fixed parameters are here chosen to provide the
BSEC existence. }}
\end{figure}

\section{Conclusion}

Some wave bending rules and some of the inverse
problem (IP) infinite classes of exactly solvable models (ESM)
[1,2] were considered. Now we have qualitative notion of {\bf
continuum spaces of ESM} instead of about 10 usually used in
contemporary quantum education. Unfortunately the majority of
physical community members is still unaware of this important
progress which  gives us unprecedented possibility to combine
enrichment of our quantum intuition with possibility to check them
numerically using complete sets of closed analytical expressions
of IP [1,2].

When considering immense number of INVERSE problem (IP)
models (from their complete sets) we began to feel
the one-to-one correspondence $S(E) --> V(x)$, it is simultaneously true in opposite
direction: in DIRECT problem (DP) $V(x)-->S(x)$.  That was one of our general goals
almost just after the beginning of our investigations to establish symmetry between two
halves of quantum knowledge, IP and DP, which was violated so strong and so long.  Really,
$S(E)$ and $V(x)$ are exactly the different representation of the same essence. Now
we often solve qualitatively many DP, using the experience of IP exact models solutions.
This can mislead the unaccustomed and not forewarned ones . What is good here is
the silent testimony of significant progress in pursuing our long dream about symmetry.
It must be also mentioned that many laws of the connections $S(E) <--> V(x)$ revealed
by the IP exact models became now qualitatively evident and enriched our quantum
intuition far beyond  the scope of exact IP results. For example, it is not always
practically important for us what occurs with the highest states (exact conservation
all spectral parameters besides the specially chosen for the transformation.
We used this, e.g., when utilizing the piecewise constant periodical potential
of Kronig-Penny for practically strict explanation the new conception of permanent
resonance which destroys the physical solutions in forbidden zones. It often appears
in small and big discoveries about what nobody has thought before, because there was no
suspect to look for something new in very familiar field. For us due to IP
approach we found
there evident now (seemingly even trivial), but previously absolutely unexpected facts.

Our opinion based on own experience about the possibility for
anyone to make discoveries \cite{5}. We consider now the
multidimensional generalization of wave bending with hope to
reveal the simplified ABC - essence of mechanism of
micro-formation of density and kinetic energy distribution between
different directions and how it depends on potential relief. It is quite different
in comparison with one dimensional case.  The new rules of wave behavior must be
in complete and instructive consistence with those for the
multichannel formalism which we have already found and considered
above.   Here is promising the local linearization of complicated
potential terrain
features allowing crucial separation of variables in nearest vicinity of each point
simplifying the seemingly inaccessible clarity of  wave control
essence.

\end{document}